# Estimating Aggregate Properties on Probabilistic Streams

Andrew McGregor        S. Muthukrishnan
andrewm@talk.ucsd.edu    muthu@google.com

**Abstract**

The probabilistic-stream model was introduced by Jayram et al. [16]. It is a generalization of the data stream model that is suited to handling "probabilistic" data where each item of the stream represents a probability distribution over a set of possible events. Therefore, a probabilistic stream determines a distribution over potentially a very large number of classical "deterministic" streams where each item is deterministically one of the domain values. The probabilistic model is applicable for not only analyzing streams where the input has uncertainties (such as sensor data streams that measure physical processes) but also where the streams are derived from the input data by post-processing, such as tagging or reconciling inconsistent and poor quality data.

We present streaming algorithms for computing commonly used aggregates on a probabilistic stream. We present the first known, one pass streaming algorithm for estimating the AVG, improving results in [16]. We present the first known streaming algorithms for estimating the number of DISTINCT items on probabilistic streams. Further, we present extensions to other aggregates such as the repeat rate, quantiles, etc. In all cases, our algorithms work with provable accuracy guarantees and within the space constraints of the data stream model.

## 1 Introduction

In order to deal with massive data sets that arrive online and have to be monitored, managed and mined in real time, the *data stream* model has become popular. In stream $A$, new item $a_t$ that arrives at time $t$ is from some universe $[n] := \{1, \ldots, n\}$. Applications need to monitor various aggregate queries such as DISTINCT, MEDIAN, HEAVY-HITTER, etc. in small space, typically in $\text{poly}(\log n)$ where poly is some polynomial. Data stream management systems are becoming mature at managing such streams and monitoring these aggregates on high speed streams such as in IP traffic analysis [9], financial and scientific data streams [3], and others. See [2, 19] for surveys of systems and algorithms for data stream management.

We go beyond this classic data stream model and consider a recently introduced generalization called the *probabilistic stream model* [16]. The central premise here is that each item in the stream is not simply one from the universe, but represents a likelihood of a set of items from the universe. Thus, the input is "probabilistic" where each item on the stream represents a probability distribution over a set of possible events. For example, consider the stream of search queries posed by users that arrives at a search engine. Based on analyzing each such query, say for example "Stunning Jaguar", one may be able to consider the probabilities associated with the different topics of interest to the user, such as "Car, Pr(Car)=0.2", "Animal, Pr(Animal)=0.7" and "Movie, Pr(Movie)=0.1". Such a probabilistic stream can be thought of as a probability distribution over potentially a very large number of classical "deterministic" streams we described first. In monitoring such a stream, the aggregates we will now be interested in are the *expected* values of the aggregates over the probabilistic streams. For example, we may be interested in the likely number of queries to the topic "Car" in the search query stream above.

The probabilistic stream model is a natural generalization of the classical data stream model and applies in a variety of ways. There are instances when the input data is probabilistic by its nature: for example, when we measure physical quantities such as light luminosity, temperature and others, we actually measure an average of these quantities over many tiny time instants in the analog world to get a single digital reading. There are instances where the input data is *uncertain* such as due to the calibration of measurement devices and the noise in measurements. There are also instances when the stream is *derived* from the input and this introduces the probabilities. For example, the original motivation for the probabilistic stream model [16] was studying uncertainties in data associated with data cleaning and reconciliation that resulted in probabilistic databases. There are also other instances when the derived data stream is probabilistic. The motivating example we provided above is such an instance in which the input is stream of web search queries and the derived stream is a stream of probability distributions over topics of queries rather than the search terms.

Previous work on probabilistic streams has included data stream algorithms for certain aggregate queries [6, 7, 16]. In this paper, we improve those results and study additional aggregate functions which are quite fundamental in any stream monitoring. In what follows, we will first describe the model precisely and then state our results.

## 1.1 The Probabilistic Stream Model and Our Problems

**Definition 1** (Probabilistic Stream). *A probabilistic stream is a data stream $A = \langle a_1, a_2, \ldots, a_m \rangle$ in which each data item $a_i$ encodes a random variable that takes a value in $[n] \cup \{\bot\}$. In particular each $a_i$ consists of a set of at most $l$ tuples of the form $(j, p_j^i)$ for some $j \in [n]$ and $p_j^i \in [0, 1]$. These tuples define the random variable $X_i$ where $X_i = j$ with probability $p_j^i$ and $X_i = \bot$ otherwise. We define $p_\bot^i = \Pr[X_i = \bot]$ and $p_j^i = 0$ if not otherwise determined.*

A probabilistic stream naturally gives rise to a distribution over "deterministic" streams. Specifically we consider the $i$th element of the stream to be determined according to the random variable $X_i$ and each element of the stream to be determined independently. Hence,

$$\Pr[\langle x_1, x_2, \ldots, x_m \rangle] = \prod_{i \in [m]} \Pr[X_i = x_i] = \prod_{i \in [m]} p_{x_i}^i .$$

We will be interested in the expected value of various quantities of the deterministic stream induced by a probabilistic stream.

**Definition 2.** *We are interested in the following aggregate properties:*

1. $\mathsf{SUM} = E\left[\sum_{i \in [m] : X_i \neq \bot} X_i\right]$

2. $\mathsf{COUNT} = E\left[|\{i \in [m] : X_i \neq \bot\}|\right]$

3. $\mathsf{AVG} = E\left[\frac{\sum_{i \in [m] : X_i \neq \bot} X_i}{|\{i \in [m] : X_i \neq \bot\}|}\right].$

4. $\mathsf{DISTINCT} = E\left[|\{j \in [n] : \exists i \in [m], \ X_i = j\}|\right]$

5. MEDIAN = $x$ such that,

$$E\left[|\{i \in [m] : X_i < x\}|\right] \leq 1/2 \lceil \text{COUNT} \rceil \text{ and } E\left[|\{i \in [m] : X_i > x\}|\right] \leq 1/2 \lceil \text{COUNT} \rceil$$

6. REPEAT-RATE $= E\left[\sum_{j \in [n]} |\{i \in [m] : X_i = j\}|^2\right]$

These are fundamental aggregates for stream monitoring, and well-studied in the classical stream model. For example, REPEAT-RATE in that case is the $F_2$ estimation problem studied in the seminal [1]. DISTINCT and MEDIAN have a long history in the classical data stream model [18, 12]. In the probabilistic stream model, we do not know of any prior algorithm with guaranteed bounds for these aggregates except the AVG (SUM and COUNT are trivial) which, as we discuss later, we will improve.

As in classical data stream algorithms, our algorithms need to use polylog($n$) space and time per item, on a single pass over the data. It is realistic to assume that $l = O(\text{polylog } n)$ so that processing each item in the stream is still within this space and time bounds. In most motivating instances, $l$ is likely to be quite small, say $O(1)$. Also, since the deterministic stream is an instance of the probabilistic stream (with each item in the probabilistic stream being one deterministic item with probability 1), known lower bounds for the deterministic stream model carry over to the probabilistic stream model. As a result, for most estimation problems, there does not exist streaming algorithms that precisely calculate the aggregate. Instead, we will focus on returning approximations to the quantities above. We say a value $\hat{Q}$ is an $(\epsilon, \delta)$-approximation for a real number $Q$ if $|\hat{Q} - Q| \leq \epsilon Q$ with probability at least $(1 - \delta)$ over its internal coin tosses. Many of our algorithms will be deterministic and return $(\epsilon, 0)$-approximations. When approximating MEDIAN it makes more sense to consider a slightly different notion of approximation. We say $x$ is an $\epsilon$-approximate median if

$$E\left[|\{i \in [m] : X_i < x\}|\right] \leq (1/2 + \epsilon) \lceil \text{COUNT} \rceil \text{ and } E\left[|\{i \in [m] : X_i > x\}|\right] \leq (1/2 + \epsilon) \lceil \text{COUNT} \rceil \text{ .}$$

**Related Models:** The probabilistic stream model complements another stream model that has been recently considered where the *stream consists of independent samples drawn from a single unknown distribution* [8, 14]. In this alternative model, the probabilistic stream $A = \langle a_1, \ldots, a_1 \rangle$ consists of a repeated element encoding a single probability distribution over $[n]$, but the crux is that the algorithm does not have access to the probabilistic stream. The challenge is to infer properties of the probability distribution $a_1$ from a randomly chosen deterministic stream. There is related work on *reconstructing strings from random traces* [5, 17]. Here, each element of probabilistic stream is of the form $\{(i, 1 - p), (\perp, p)\}$ for some $i \in [n]$. As before, the algorithm does not have access to the probabilistic stream but rather tries to infer the probabilistic stream from a limited number of independently generated deterministic streams. The results in [14, 8, 5, 17] do not provide any bounds for estimating aggregates such as the ones we study here when input is drawn from multiple, known probability distributions.

### 1.2 Our Results

We present the first single-pass algorithms for estimating the aggregate properties AVG, MEDIAN, REPEAT-RATE, and DISTINCT. The algorithms for AVG and MEDIAN are deterministic while the other two algorithms are randomized. While it is desirable for all the algorithms to be deterministic, this randomization can be shown to be necessary using the standard results in streaming algorithms, eg., of Alon, Matias, and Szegedy [1, Proposition 3.7].

Throughout this paper we assume each probability specified in the probability stream is either 0 or $\Omega(1/\operatorname{poly}(n))$. The space-complexity of our algorithms are as follows.

- We present a single pass, deterministic, $(\epsilon, 0)$-approximation for AVG using $O(\epsilon^{-1} \log(nm/\epsilon))$ space. Further, if the COUNT of the stream is sufficiently high then our algorithm needs only $O(1)$ words of space.

  The best known previous work [16] presents an $O(\log n)$ pass algorithm using $O(\epsilon^{-1} \log^2 n)$ space. Thus our algorithm substantially improves the number of passes needed and indeed works in the one-pass constraint of a streaming model.

- We present the first known streaming algorithms for other aggregates listed earlier. The rest of our results are summarized in the table below.

|  | Space | Randomized/Deterministic |
|---|---|---|
| DISTINCT | $O(\epsilon^{-5} \log n \log^2 \delta^{-1})$ | Randomized |
| REPEAT-RATE | $O(\epsilon^{-2}(\log n + \log m) \log \delta^{-1})$ | Randomized |
| MEDIAN | $O(\epsilon^{-2} \log m)$ | Deterministic |

The aggregates such as AVG, SUM and COUNT appear to be trivial to estimate, at first glance. That is the case with deterministic streams. For randomized streams, SUM and COUNT can still be estimated easily since they are linear in the input and hence we can write straightforward formulas to compute them. However AVG is not simply SUM/COUNT as shown in [16] and needs nontrivial techniques. The central difficulty is that it is nonlinear and hence the principle of "linearity of expectation" is not directly useful. The other aggregates such as DISTINCT, MEDIAN and REPEAT-RATE have previously known algorithms for deterministic streams. A natural approach therefore would be to randomly instantiate multiple independent deterministic streams, apply standard stream algorithms and then apply a robust estimator. This approach works to an extent but typically gives poor approximations to small quantities. This is the case for DISTINCT. Furthermore, this approach necessitates a high degree of randomization. For MEDIAN and REPEAT-RATE, we show that it is possible to deterministically instantiate a single deterministic stream and run standard stream algorithms.

## 2   Average, Sum, and Count

For the duration of this section let $Y$ and $Z$ be the random variables defined by,

$$Y = \sum_{i: X_i \neq \perp} X_i \quad \text{and} \quad Z = |\{i \in [m] : X_i \neq \perp\}| \ .$$

In this section we are interested in estimating $\mathsf{AVG} = E[Y/Z]$. It was shown in [16] that $E[Y]/E[Z]$ can be an arbitrarily poor approximation for AVG.

We consider two different cases. We show that if stream has a sufficiently long expected length, that is, COUNT is large, then $E[Y]/E[Z]$ is a guaranteed approximation for AVG. This will follow because $Z$ will be tightly concentrated around $E[Z]$. Subsequently we show AVG can be approximated for streams with shorter expected length. This needs a different idea: if COUNT is not large then it is sufficient to estimate AVG from $\Pr[Z = z]$ and $E[Y|Z = z]$ for a relatively small range of values of $z$. When combined, this gives us a single pass algorithm improving upon the algorithm presented in [16] that uses $O(\log n)$ passes.

## 2.1 Streams with long expected length

**Lemma 1.** *If* $\mathsf{COUNT} \geq 12\epsilon^{-2}\ln(10nm\epsilon^{-1})$ *then,*

$$\left|\frac{\mathsf{SUM}}{\mathsf{COUNT}} - \mathsf{AVG}\right| \leq \epsilon \mathsf{AVG} \ .$$

*Proof.* Let $A$ be the event that $|Z - \mathsf{COUNT}| \geq \epsilon \mathsf{COUNT}/2$. Assume that $\mathsf{COUNT} \geq 12\epsilon^{-2}\ln(2\delta^{-1})$ and so, by an application of the Chernoff-Hoeffding bounds,

$$\Pr[A] \leq 2\exp(-\epsilon^2 \mathsf{COUNT}/12) \leq \delta \ .$$

Set $\delta = \epsilon/(5nm)$. Then,

$$\begin{aligned}
\left|\frac{\mathsf{SUM}}{\mathsf{COUNT}} - \mathsf{AVG}\right| &= \left|\frac{E[Y]}{E[Z]} - \sum_{y,z}\frac{y}{z}\Pr[Y=y, Z=z, A] - \sum_{y,z}\frac{y}{z}\Pr[Y=y, Z=z, \neg A]\right| \\
&\leq \left|\frac{E[Y]}{E[Z]} - \sum_{y,z}\frac{y}{z}\Pr[Y=y, Z=z, \neg A]\right| + \sum_{y,z}\frac{y}{z}\Pr[Y=y, Z=z, A] \\
&\leq \left|\frac{E[Y]}{E[Z]} - \frac{1}{E[Z]}\sum_y y\Pr[Y=y, \neg A]\right| + \frac{\epsilon}{2E[Z]}\sum_y y\Pr[Y=y, \neg A] + n\Pr[A] \\
&\leq \left|\frac{E[Y]}{E[Z]} - \frac{1}{E[Z]}\sum_y y\Pr[Y=y, \neg A]\right| + \frac{\epsilon E[Y]}{2E[Z]} + n\delta \\
&\leq \frac{\epsilon E[Y]}{2E[Z]} + nm\delta + n\delta \\
&\leq \frac{9\epsilon}{10}\frac{E[Y]}{E[Z]} \ .
\end{aligned}$$

where the last line follows since $E[Y] \geq E[Z]$. Consequently, for sufficiently small $\epsilon$, we deduce that $\left|\frac{\mathsf{SUM}}{\mathsf{COUNT}} - \mathsf{AVG}\right| \leq \epsilon \mathsf{AVG}$ as required. □

COUNT and SUM can be trivially computed exactly in a single pass because, $\mathsf{COUNT} = \sum_{i\in[m]}(1-p_\perp^i)$ and $\mathsf{SUM} = \sum_{i\in[m]} E[X_i|X_i \neq \perp](1-p_\perp^i)$. This leads to the following theorem.

**Theorem 1.** *If* $\mathsf{COUNT} \geq 12\epsilon^{-2}\ln(10nm\epsilon^{-1})$ *then we can deterministically estimate* AVG *in a single pass using* $O(\log n)$ *space.*

## 2.2 The General Case

In this section we will remove the assumption that $\mathsf{COUNT} \geq 12\epsilon^{-2}\ln(10nm\epsilon^{-1})$ but our algorithm will require more space. The main idea behind our algorithm is that if COUNT is not large then it is sufficient to estimate AVG from $\Pr[Z = z]$ and $E[Y|Z = z]$ for a relatively small range of values of $z$.

The next lemma shows that it is possible compute the probability that $Z$ takes various values if we assume that at each point during the evolution of the stream, the number of elements that have appeared does not differ significantly from its expected number.

**Lemma 2.** *For $j \in [m]$, let $Z_j = |\{i \in [j] : X_i \neq \perp\}|$ and $Y_j = \sum_{i \leq j: X_i \neq \perp} X_i$. Let $C_j^w$ be the event that,*
$$\forall i \in [j], \ |Z_i - E[Z_i]| \leq w \ .$$
*It is possible to compute $A_z = \Pr[Z = z, C_m^w]$ and $B_z = \sum_y y \Pr[Y = y, Z = z, C_m^w]$ for all $z$ in a single pass using $O(w \log n)$ bits of space.*

*Proof.* Let $A_{j,z} = \Pr\left[Z_j = z, C_j^w\right]$ and $B_{j,z} = \sum_y y \Pr\left[Y_j = y, Z_j = z, C_j^w\right]$. First note that for $j, z \in [m]$,

$$\begin{aligned}
A_{j,z} &= \Pr\left[Z_{j-1} = z, X_j = \perp, C_j^w\right] + \Pr\left[Z_{j-1} = z-1, X_j \neq \perp, C_j^w\right] \\
&= \begin{cases} A_{j-1,z-1}(1 - p_\perp^j) + A_{j-1,z} p_\perp^j & \text{if } |z - E[Z_j]| \leq w \\ 0 & \text{otherwise} \end{cases}
\end{aligned}$$

where $A_{0,z} = 1$ if $z = 0$ and 0 otherwise. Also, on the assumption that $|z - E[Z_j]| \leq w$,

$$\begin{aligned}
B_{j,z} &= \sum_y y \left( \Pr\left[Y_{j-1} = y, Z_{j-1} = z, X_j = \perp, C_j^w\right] \right. \\
&\quad \left. + \sum_{a \in [n]} \Pr\left[Y_{j-1} = y - a, Z_{j-1} = z-1, X_j = a, C_j^w\right] \right) \\
&= p_\perp^j B_{j-1,z} + (1 - p_\perp^j) \sum_a \Pr[X_j = a | X_j \neq \perp] \sum_y y \Pr\left[Y_{j-1} = y - a, Z_{j-1} = z-1, C_j^w\right] \\
&= p_\perp^j B_{j-1,z} + (1 - p_\perp^j) \sum_a \Pr[X_j = a | X_j \neq \perp] (a \Pr\left[Z_{j-1} = z-1, C_j^w\right] + B_{j-1,z-1}) \\
&= p_\perp^j B_{j-1,z} + (1 - p_\perp^j)(E[X_j | X_j \neq \perp] A_{j-1,z-1} + B_{j-1,z-1})
\end{aligned}$$

and $B_{0,z} = 0$ for all $z$. If $|z - E[Z_j]| > w$ then clearly, $B_{j,z} = 0$. Lastly $E[Z_j] = E[Z_{j-1}] + (1 - p_\perp^j)$. Hence we can compute $(A_{j,z})_{0 \leq z \leq m}$, $(B_{j,z})_{0 \leq z \leq m}$ and $E[Z_j]$ from $(A_{j-1,z})_{0 \leq z \leq m}$, $(B_{j-1,z})_{0 \leq z \leq m}$ and $E[Z_{j-1}]$. The space bound follows because at most $O(w)$ of $(A_{j,z})_{0 \leq z \leq m}$ and $(B_{j,z})_{0 \leq z \leq m}$ are non-zero for any $j$. □

Hence, if we choose $w$ to be large enough such that $C_m^w$ is a sufficiently high probability event or that $\mathsf{COUNT} \geq 12\epsilon^{-2} \ln(10nm\epsilon^{-1})$, then we can use the above lemma to obtain a deterministic algorithm for estimating AVG. We show:

**Theorem 2.** *AVG can be computed to $(\epsilon, 0)$ approximation in a single pass using $O(\epsilon^{-1}(\log n + \log m + \log \epsilon^{-1}) \log n)$ bits of space.*

*Proof.* Let $c = 12\epsilon^{-2}\ln(10nm\epsilon^{-1})$. First, if COUNT $\geq c$ then by Theorem 1, SUM/COUNT is an $(\epsilon, 0)$ approximation of AVG. Alternatively assume that COUNT $\leq c$. Let $w = \epsilon c$. By an application of the Chernoff-Hoeffding and union bounds,

$$\Pr[C_m^w] \geq 1 - \sum_{i \in [m]} \Pr[|Z_i - E[Z_i]| > w] \geq 1 - 2m\exp(-\epsilon^2 c/3) \geq 1 - \epsilon/4n$$

$\sum_z \frac{1}{z} \sum_y y \Pr[Y = y, Z = z, C_m^w]$ can be computed in $O(w \log n)$ space as described in Lemma 2. But then,

$$\left| \sum_z \frac{1}{z} \sum_y y \Pr[Y = y, Z = z, C_m^w] - \mathsf{AVG} \right| = \sum_{y,z} \frac{y}{z} \left| \Pr[Y = y, Z = z, C_m^w] - \Pr[Y = y, Z = z] \right|$$
$$\leq n \Pr[\neg C_m^w]$$
$$\leq \epsilon .$$

This translates into a $(\epsilon, 0)$ approximation since AVG $\geq 1$. □

**Precision Issues:** One remaining issue that needs to be addressed is the precision to which the various quantities are calculated. In Lemma 2, it was assumed that all the arithmetic could be performed exactly. However, it is very likely that, for some $z, j$ both $A_{j,z}$ and $B_{j,z}$ can be exponentially small and yet non-zero. We can easily show that errors introduced by a lack of precision do not accumulate sufficiently to cause a problem. In particular, let us assume that the calculation of each $A_{j,z}$ is performed with the addition of error $\beta$. Then, given that $A_{j,z} = A_{j-1,z-1}(1 - p_\perp^j) + A_{j-1,z}p_\perp^j$, if it is non-zero, a simple induction argument shows that $A_{j,z}$ is computed with error at most

$$(p_\perp^j + 1 - p_\perp^j)(j - 1)\beta + \beta = j\beta .$$

Similarly, $B_{j,z}$ is computed with error at most,

$$p_\perp^j n(j - 1)^2 \beta + (1 - p_\perp^j)(n(j - 1)\beta + n(j - 1)^2 \beta) + \beta \leq j^2 n\beta .$$

Therefore $\beta = \epsilon m^{-2} n^{-2}$ suffices to compute $\sum_{y,z} \frac{y}{z} \Pr[Y = y, Z = z, C_m^w]$ with additive-error at most $\epsilon$ which translates into a relative-error since the quantity being estimated is $\Omega(1)$.

## 3 Distinct Items

In this section we present a $(\epsilon, \delta)$-approximation algorithm for DISTINCT. In contrast to the algorithm for AVG, a major part of our algorithm is to actually randomly instantiate numerous deterministic streams and to compute the average value of the number of distinct values in these streams. However, this general approach will only give an $(\epsilon, \delta)$ approximation in small space if the expected number of distinct values is not very small. In the following theorem we show that it is possible to also deal with this case. Furthermore, the random instantiations will be performed in a slightly non-obvious fashion for the purpose of achieving a tight analysis of the appropriate probability bounds.

**Theorem 3.** *We can $(\epsilon, \delta)$-approximate DISTINCT in one pass and $O(\epsilon^{-5} \log n \log^2 \delta^{-1})$ space.*

*Proof.* First note that,

$$\mathsf{DISTINCT} = \sum_{j \in [n]} \Pr\left[\exists i \in [m], \ X_i = j\right] = \sum_{j \in [n]} \left(1 - \prod_{i \in [m]} (1 - p_j^i)\right) \ .$$

Consider $\mathsf{COUNT} = \sum_i (1 - p_\perp^i)$. Then,

$$e^{-\mathsf{COUNT}} \mathsf{COUNT} \leq 1 - e^{-\mathsf{COUNT}} \leq 1 - \prod_{i \in [m], j \in [n]} (1 - p_j^i) \leq \mathsf{DISTINCT} \leq \sum_{i \in [m], j \in [n]} p_j^i = \mathsf{COUNT} \ .$$

Hence if $\mathsf{COUNT} \leq \ln(1+\epsilon)$ then $\mathsf{COUNT}$ is an $(\epsilon, 0)$ approximation for $\mathsf{DISTINCT}$. We now assume that $\mathsf{COUNT} > \ln(1+\epsilon)$ and hence

$$\mathsf{DISTINCT} \geq \ln(1+\epsilon) e^{-\ln(1+\epsilon)} \geq \epsilon/2$$

assuming $\epsilon$ is sufficiently small.

Consider the following basic estimator $X$,

1. For each tuple $(j, p_j^i)$ in $a_i$, put $j$ in an induced stream $A'$ with probability $p_j^i$

2. Compute an $(\epsilon/3, \delta/(2c_1))$ approximation $X$ of $F_0(A')$ using [4].

Note that the stream $A'$ is not generated according to the distribution defined by the probabilistic stream because more than one item can be generated for a given $a_i$. However, the expected number of distinct elements in $A'$ is the same as the expected number of distinct elements in a stream generated by the probabilistic stream $A$. In particular $E[X] = (1 \pm \epsilon/3)\mathsf{DISTINCT}$. The reason for generating $A'$ in this way is that we may argue that the events that $i$ and $j$ appear in $A'$ are independent. This will be important in our analysis.

We will compute $c_1 = 3^3 \cdot 2\epsilon^{-3} \ln(4/\delta)$ of these basic estimators and average them to produce our final estimator $Y$. The accuracy and probability of success of our algorithm follows from the following claim.

**Claim 3.** $\Pr[|Y - \mathsf{DISTINCT}| \leq \epsilon \mathsf{DISTINCT}] \leq \delta$

*Proof.* First we effectively assume that $F_0(A')$ can be computed exactly. Let $Z$ be the random variable corresponding to the average of $c_1$ copies of $F_0(A')$. $E[Z] = \mathsf{DISTINCT}$. Note that $c_1 Z$ can be thought of as the sum of $nc_1$ independent boolean trials: $c_1$ trials corresponding to each $j \in [n]$ (some trials may have zero probability of success) and hence,

$$\begin{aligned}
\Pr[|Z - \mathsf{DISTINCT}| \leq (1 + \epsilon/3)\mathsf{DISTINCT}] &= \Pr[|c_1 Z - c_1 \mathsf{DISTINCT}| \leq (1 + \epsilon/3) c_1 \mathsf{DISTINCT}] \\
&\leq 2 \exp(-\epsilon^2 c_1 \mathsf{DISTINCT}/27) \\
&\leq 2 \exp(-\epsilon^3 c_1/(2 \cdot 27)) \\
&\leq \delta/2 \ .
\end{aligned}$$

With probability at least $1 - c_1 \delta/(2c_1) = 1 - \delta/2$ each of the $c_1$ calls to the distinct element counter returns an $\epsilon/3$ approximation. Hence with probability at least $1 - \delta/2$, $Y = (1 \pm \epsilon/3)Z$. Therefore with probability at least $1 - \delta$,

$$Y = (1 \pm \epsilon/3)Z = (1 \pm \epsilon/3)^2 \mathsf{DISTINCT} \leq (1 \pm \epsilon) \mathsf{DISTINCT}$$

assuming $\epsilon < 1/4$. □

The space bound follows because for each of our $O(\epsilon^{-3} \log \delta^{-1})$ basic estimators, the distinct element counter requires $O((\epsilon^{-2} + \log n) \log \delta^{-1})$ space [4]. □

## 4 Other Aggregates

In this section, we present streaming algorithms for other aggregates such as REPEAT-RATE and MEDIAN. In both cases, the approach is to reduce the problem to ones over the deterministic streams, and the algorithms are pleasantly simple.

**Estimating REPEAT-RATE:** We present an $(\epsilon, \delta)$-approximation for REPEAT-RATE. The following result is based on reducing the problem to estimating $F_2$ in a deterministic stream. In contrast to the result on estimating DISTINCT, this reduction can be done deterministically. We then use an algorithm by Alon, Matias, and Szegedy [1] to estimate the second frequency moment of the deterministic stream.

**Theorem 4.** *There exists a single pass algorithm that $(\epsilon, \delta)$-approximates REPEAT-RATE using $O(\epsilon^{-2}(\log m + \log n) \log \delta^{-1})$ bits of space.*

*Proof.* We start by re-writing REPEAT-RATE as follows.

$$
\begin{aligned}
\mathsf{REPEAT\text{-}RATE} &= \sum_{j \in [n]} E\left[|\{i : X_i = j\}|^2\right] \\
&= \sum_{j \in [n]} \left( \sum_{i,k \in [m]: i \neq k} p_j^i p_j^k + \sum_{i \in [m]} p_j^i \right) \\
&= \sum_{j \in [n]} \left( \sum_{i \in [m]} p_i^j \right)^2 + \sum_{i \in [m]} p_i^j (1 - p_i^j)
\end{aligned}
$$

Note that this first term can be $(\epsilon, \delta)$-approximated using $O(\epsilon^{-2} \log \delta^{-1}(\log m + \log n))$ space using an algorithm of Alon, Matias, and Szegedy [1]. The second term can be computed exactly in $O(\log m)$ space. Since both terms are positive we get an $(\epsilon, \delta)$-approximation to REPEAT-RATE in the space claimed. □

**Estimating MEDIAN:** We present an algorithm for finding an $\epsilon$-approximate MEDIAN. Again our result is based on a deterministic reduction of the problem to median finding in a deterministic stream. To solve this problem we use the algorithm of Greenwald and Khanna [13].

**Theorem 5.** *There exists a single pass algorithm finding an $\epsilon$-approximate MEDIAN in $O(\epsilon^{-1} \log m)$ space.*

*Proof.* The idea is use the selection algorithm of Greenwald, Khanna [13], on an induced stream as follows.

1. For each tuple $(j, p_j^i)$ in $a_i$, put $\left\lfloor 2m p_j^i \epsilon^{-1} \right\rfloor$ copies of $j$ in an induced stream $A'$.

2. Using the algorithm of Greenwald, Khanna [13], find an element $l$ such that

$$|\{i \in A : 1 \leq i < l\}| \leq (1/2 + \epsilon/2)|A'| \text{ and } |\{i \in A : l < i \leq n\}| \leq (1/2 + \epsilon/2)|A'| \ . \qquad (1)$$

where $|A'|$ denotes the length of the stream $A'$.

Note that $p_j^i \geq \frac{\lfloor 2mp_j^i\epsilon^{-1} \rfloor}{2m\epsilon^{-1}} \geq p_j^i - \epsilon/(2m)$. Therefore, dividing Eq. 1 by $2m\epsilon^{-1}$ yields

$$\left( \sum_{1 \leq j < l, i \in [m]} p_j^i \right) - \epsilon/2 \leq (1/2+\epsilon/2) \lceil \mathsf{COUNT} \rceil \text{ and } \left( \sum_{l < j \leq n, i \in [m]} p_j^i \right) - \epsilon/2 \leq (1/2+\epsilon/2) \lceil \mathsf{COUNT} \rceil \ .$$

The result follows since $\mathsf{COUNT} > 0$. □

## 5 Concluding Remarks

A number of remarkable algorithmic ideas have been developed for estimating aggregates over deterministic streams since the seminal work of [1]. Some of them are applicable to estimating aggregates over probabilistic streams such as in estimating MEDIAN and REPEAT-RATE by suitable reductions, but for other aggregates such as DISTINCT and AVG, we need new ideas that we have presented here. The probabilistic stream model was introduced in [16] mainly motivated by probabilistic databases where data items have a distribution associated with them because of the uncertainties and inconsistencies in the data sources. This model has other applications too, including in the motivating scenario we described here in which the stream (topic distribution of search queries) derived from the deterministic input stream (search terms) is probabilistic. We believe that the probabilistic stream model will be very useful in practice in dealing with such applications.

There are several technical and conceptual open problems. For example, could one characterize problems for which there is a (deterministic or randomized) reduction from probabilistic streams to deterministic streams without significant loss in space bounds or approximations? We suspect that for additive approximations, there is a simple characterization. Also, can we extend the solutions for estimating the basic aggregates we have presented here to others, in particular, geometric aggregates [15] or aggregate properties of graphs [10, 11]?